\documentclass[submission,copyright,creativecommons]{eptcs}
\providecommand{\event}{ThEdu 2024} 

\usepackage{iftex}

\ifpdf
  \usepackage{underscore}         
  \usepackage[T1]{fontenc}        
\else
  \usepackage{breakurl}           
\fi

\usepackage{xspace}
\usepackage{graphicx} 
\usepackage{caption}  
\usepackage{float}    
\usepackage{wrapfig}
\usepackage{cite}
\usepackage{tikz}
\usepackage{todonotes}
\usetikzlibrary{shapes, arrows, arrows.meta, positioning, decorations.pathreplacing, automata}
\def\onlineprover{\textsf{OnlineProver}\xspace}

\newcommand{\titlerunning}{OnlineProver: Experience with a Visualisation Tool for Teaching Formal Proofs}
\title{\titlerunning}

\author{
Ján Perháč
\thanks{\vspace{-4mm}
\begin{wrapfigure}{l}{.2\textwidth}\vspace{-4mm}
  \includegraphics[width=.2\textwidth]{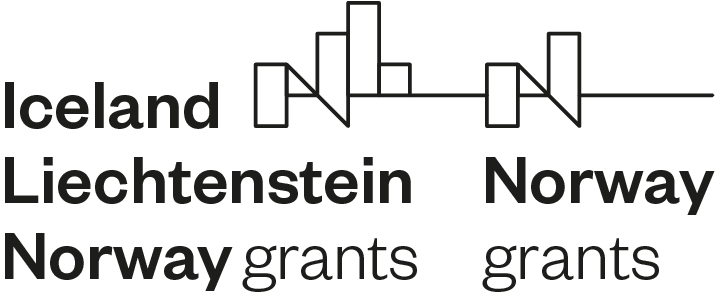}
\end{wrapfigure}
\noindent The authors thanks EEA and Norway Grants for support of this work under initiative no. FBR-PDI-025.\\ ``\emph{Working together for a green, competitive and inclusive Europe.}''
\\ \url{https://www.eeagrants.sk/}
}%
~
\thanks{The author thanks the ``European Research Network on Formal Proofs'' under EU COST Action CA20111 for support of this work.}
 \qquad\qquad
Samuel Novotný \qquad\qquad
Sergej Chodarev
\institute{
  Department of Computers and Informatics,
  Faculty of Electrical Engineering and Informatics,\\
    Technical University of Košice,
    Košice, Slovakia}
   \email{\{jan.perhac,samuel.novotny,sergej.chodarev\}@tuke.sk}
\and
Joachim Tilsted Kristensen
 \qquad\qquad
Lars Tveito
\institute{
  Department of Informatics,
  University of Oslo,
  Oslo, Norway}
  \email{\{joachkr,larstvei\}@ifi.uio.no}
\and
Oleks Shturmov \qquad\qquad
Michael Kirkedal Thomsen
\institute{
  Department of Informatics,
  University of Oslo,
  Oslo, Norway}
\institute{
  Department of Computer Science,
  University of Copenhagen,
  Copenhagen, Denmark
}
  \email{\{michakt,olekss\}@ifi.uio.no}
}

\newcommand{\authorrunning}{J. Perháč et al.}

\hypersetup{
  bookmarksnumbered,
  pdftitle    = {\titlerunning},
  pdfauthor   = {\authorrunning},
  pdfsubject  = {EPTCS},               
}

\begin{document}
\maketitle

\begin{abstract}
\onlineprover is an interactive proof assistant tailored for the educational setting. Its main features include a user-friendly interface for editing and checking proofs. The user interface provides feedback directly within the derivation, based on error messages from a proof-checking web service. A basic philosophy of the tool is that it should aid the student while still ensuring that the students construct the proofs as if they were working on paper.

We gathered feedback on the tool through a questionnaire, and we conducted an intervention to assess its effectiveness for students in a classroom setting, alongside an evaluation of technical aspects. The initial intervention showed that students were satisfied with using \onlineprover as part of their coursework, providing initial confirmation of the learning approach behind it. This gives clear directions for future developments, with the potential to find and evaluate how \onlineprover can improve the teaching of natural deduction.
\end{abstract}

\section{Introduction}
\label{sect:introduction}

When teaching theoretical computer science courses (e.g., introduction to first-order logic), the focus is often not on programming; instead, we aim to teach an understanding of discrete abstractions and their properties through proofs. For this reason, spending several weeks of the course on introducing a theorem prover can be counterproductive, as it diverts the attention of students from the foundational concepts. In our project\footnote{\url{https://onlineprover.github.io/}}, we propose a tool, \onlineprover\footnote{\url{http://onlineprover.com/}}, which provides exercises for Gentzen-style derivation proofs over user-definable deduction systems.

The main objective is to build an educational tool that include some of the capabilities of computer-aided theorem proving, without requiring students to learn a new syntax or dedicating several lectures to introducing the tools. Our initial setup involves a tool for which the teacher must learn a simple domain-specific language (DSL) for specifying derivation systems and exercises. However, students use the tool solely to solve exercises provided by the teacher through a web interface designed for this purpose. The interface is capable of checking students’ derivations and providing feedback directly within the derivation where it is needed.

Computer science (CS) students typically follow several programming courses before delving into theoretical computer science, during which they gain extensive experience through a trial-and-error approach~\cite{weintropWilensky2015}. On the other hand, theoretical computer science courses all require mathematical prerequisites, which CS students frequently struggle with. Initial research~\cite{morgan} and personal experiences~\cite{newsome2023mathematical} suggest that the pedagogical approaches effective for teaching CS students differ from those tailored to mathematics students. Similar findings are observed in the field of physics education~\cite{sherin2001}.

Building upon this foundation, we have developed \textit{onlineprover}, a tool that aids CS students by facilitating the exercise of mathematical rigour while offering a taste of trial-and-error exploration. Students must specify all derivations and type all rule constructs of the proofs; there is no interactive rule selection. We want the tool to closely resemble classroom teaching with pen and paper. The learning objectives that \onlineprover could support in a course include:
\begin{itemize}
  \item Skills in deciding and proving formal properties of logical formulas (e.g., satisfiability, validity, implication, and equivalence) through natural deduction arguments.
\end{itemize}

Following the tradition in CS education~\cite{edwards2004}, we therefore aim to foster deep reflection through meticulous tool design, rather than solely promoting problem-solving through trial-and-error. Drawing from experiences in introductory programming, we acknowledge that while students appreciate the simplicity of block-based frameworks (e.g., Scratch), they may find them lacking in authenticity~\cite{weintropWilensky2015}. Research on tools for automated assessment (auto-graders)~\cite{deebHickey2023,karavirtaEtal2006,baniassadEtal2021}, often used for giving quick and direct feedback on programming tasks, has also shown that this can shift the students' focus away from problem-solving. Instead, students tend to optimize their efforts towards passing specific tests.

In this work, we have introduced the \onlineprover in at the course
``Foundations of Computing - Discrete Mathematics''~\cite{itucoursefoundations2024} with 165 students, at the IT University of Copenhagen, Denmark. A first and simple intervention were presented in~\cite{KristensenEtal:2024}
The following is a description of our first
experience use of mechanical reasoning systems for teaching. First, we will give the research questions on the work (Section~\ref{sec:reseachquestion}) followed by an outline for related work (Section~\ref{sect:relatedwork}). Then, in
Section~\ref{sec:tooldesign} we outline the design of \onlineprover, after
which Section~\ref{sect:Teaching} will exemplify the experience of the
students. In Section~\ref{sect:evaluation} we present the intervention and
the results. These will be discussed in Section~\ref{sec:discus} and finally
in Section~\ref{sect:conclusion} we conclude.
The online version of \onlineprover tool along with all exercises that is used in this experiment, can be found at \url{http://onlineprover.com/}.


\section{Research Questions}
\label{sec:reseachquestion}

To guide our research, we formulated the following research questions focusing on the usage, strategies, and perceived effectiveness of \onlineprover as a teaching and learning tool for formal proofs.
\begin{itemize}
	\item RQ1: Is our tool used as a proof assistant or a proof checker?

	\item RQ2: How do students use trial-and-error strategies with \onlineprover to overcome difficulties in solving proofs? 

	\item RQ3: Do students perceive the tool’s feedback useful?

	\item RQ4: Do students consider learning proofs using OnlineProver more effective than learning proofs by pen and paper?
\end{itemize}
We will answer these questions in Section \ref{sec:discus}.

\section{Related Work}
\label{sect:relatedwork}

Although traditional proof assistants such as Coq, Agda, and Isabelle use a programming style, we do not find them well-suited for introductory courses. It takes a significant amount of time for students to learn the underlying system's languages to be able to start working on formal proofs.
Several other tools have been developed with a focus on a domain-specific framework and implement a trial-and-error approach, giving students less room for reflection. Such tools are dedicated to various frameworks, such as proof trees for Hoare logic~\cite{korkut2023}, sequent calculus~\cite{ehle2018}, Gentzen-style proofs~\cite{gasquetEtal2011panda}, Tableau~\cite{vasconcelos2023, virsedaEtal2011}, and $\lambda$-calculus~\cite{frank2018}. While these tools often include automation and rule limitations, they are generally designed to encourage trial-and-error, which can diminish the opportunity for deep reflection.

For example, the work by~\cite{korkut2023} presents an online tool for Hoare logic that includes LaTeX export, as well as automation and rule limitations, generally promoting a trial-and-error approach. Similarly, \cite{machin2011yoda, seligman2015teaching, broda2007} discuss tools relevant to natural deduction, though some lack comprehensive references. In the context of sequent calculus, \cite{ehle2018} provides a well-written paper with a focus on first-order logic. The Gentzen-style approach is discussed in \cite{gasquetEtal2011panda}, while Tableau methods are explored by \cite{vasconcelos2023} and \cite{virsedaEtal2011}. Lastly, the $\lambda$-calculus framework is presented in \cite{frank2018}.
However, these tools are often designed to be limiting for students, exhibiting similar issues to block-based programming frameworks. They might simplify initial learning but at the cost of limiting deeper understanding and flexibility.
Closer to our approach is the work by \cite{leach-krouse2018}, which in a more general and flexible framework implements support for Fitch-style proofs. Also relevant is \cite{zenker2011}, which, though it focuses on elementary symbolic logic, takes a teaching approach aligned with our educational goals.




\section{Tool Design}
\label{sec:tooldesign}

\begin{figure}[ht!]
  \centering
  \begin{tikzpicture}[>=stealth, shorten >= 5pt, node distance=1em, scale=1]
    \pgfdeclarelayer{bg}    
    \pgfsetlayers{bg,main}  
    \tikzstyle{vertex} = [circle, scale=0.5, fill=black!10, draw=black]
    \tikzset{every arrow/.style={Stealth[scale=1.2]}} 
    \tikzstyle{op} = [midway, above=-3pt, sloped, text=black, font=\footnotesize]

    \node (editor0) at (0, 0) {Editor};
    \node (server0) at (6, 0) {Server};
    \node (engine0) at (8, 0) {Proof engine};

    \node[vertex, below = 1em of editor0] (editor1) {};
    \node[below = 2em of editor0] (editor2) {\includegraphics[height=2em]{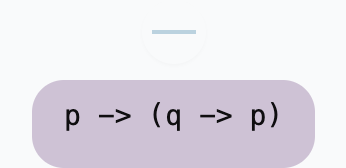}};
    \node[below = 6em of editor0] (editor3) {\includegraphics[height=2em]{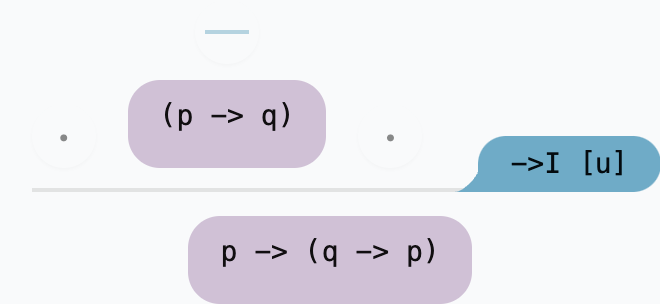}};
    \node[below = 10em of editor0] (editor4) {\includegraphics[height=2em]{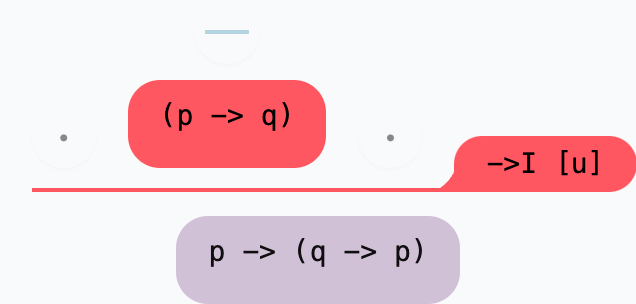}};
    \node[below = 14em of editor0] (editor5) {\includegraphics[height=2em]{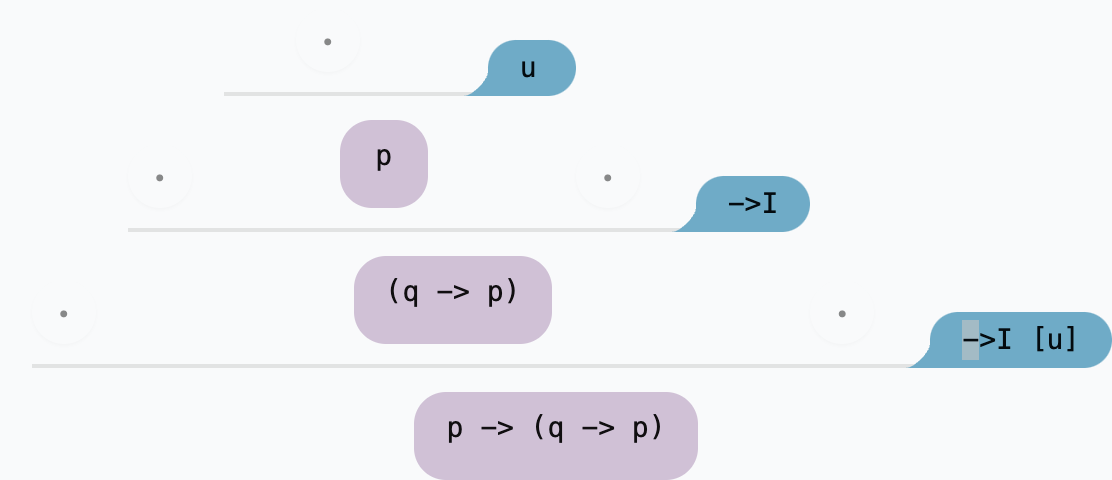}};
    \node[below = 18em of editor0] (editor6) {\includegraphics[height=2em]{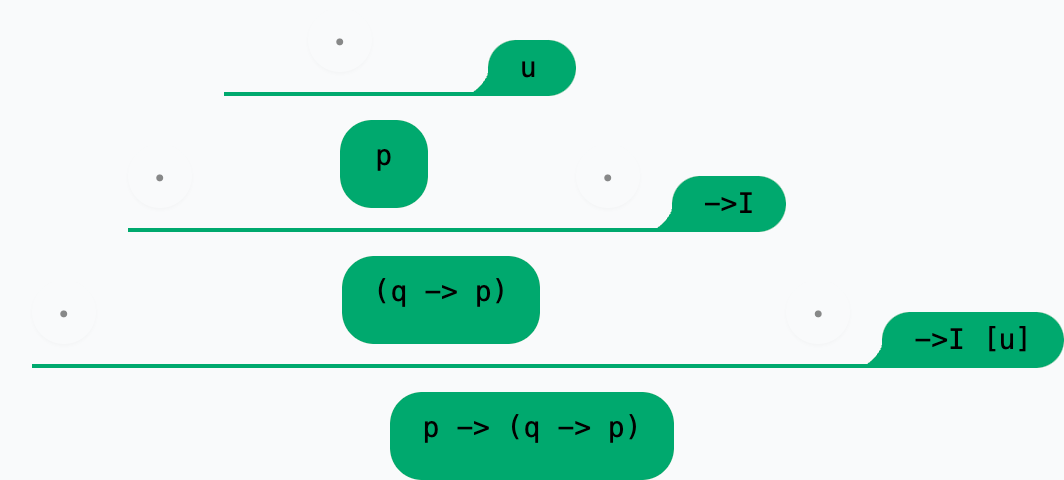}};

    \node[below = 2em of server0] (server1) [vertex] {};
    \node[below = 9em of server0] (server2) [vertex] {};
    \node[below = 17em of server0] (server3) [vertex] {};

    \node[below = 9em of engine0] (engine1) [vertex] {};
    \node[below = 17em of engine0] (engine2) [vertex] {};


    \begin{pgfonlayer}{bg}
      \draw[->, color=black!30] (editor0) -- (0, -23em);
      \draw[->, color=black!30] (server0) -- (6, -23em);
      \draw[->, color=black!30] (engine0) -- (8, -23em);
    \end{pgfonlayer}

    \draw[->] (editor1) -- (server1) node[op] {request exercise};
    \draw[->] (server1) -- (editor2) node[op] {};

    \draw[->] (editor3) -- (server2) node[op] {check};
    \draw[<->, shorten >= 5pt, shorten <= 5pt] (server2) -- (engine1);
    \draw[->] (server2) -- (editor4) node[op] {};

    \draw[->] (editor5) -- (server3) node[op] {check};
    \draw[<->, shorten >= 5pt, shorten <= 5pt] (server3) -- (engine2);
    \draw[->] (server3) -- (editor6) node[op] {};

  \end{tikzpicture}
  \caption{An example of interacting with \onlineprover.}
  \label{fig:onlineprover-protocol}
\end{figure}
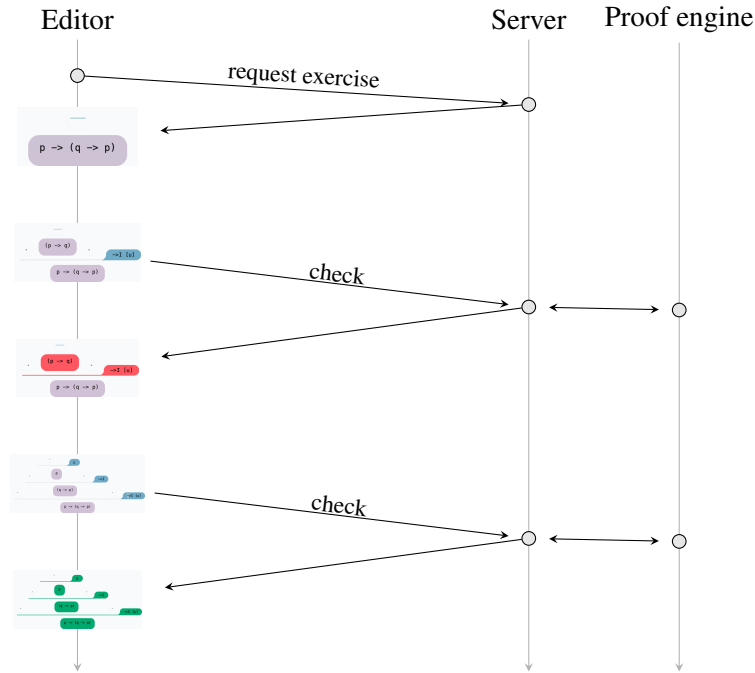

\onlineprover consists of three modular components: a proof engine, a web
server, and a web-based editor for Gentzen-style derivation trees. The proof
engine checks derivations, provides feedback, and supports alternative calculi
implementations. In this paper, the
proof engine used is based on natural deduction from \emph{The Logic Manual} by
Volker Halbach~\cite{halbach:2010}. The web server acts as an intermediary
between the user-facing editor and the proof engine, providing endpoints for
serving a front page, a collection of exercises, and requesting feedback on
exercises. The editor allows users to select exercises, where each exercise
contains a description of the task and along with one or more editable proof
trees that comprise their answer. Derivation trees are edited by adding or
deleting nodes, where each node is a plain text field. Exercises can be
checked, which submits the derivations of the exercise to be annotated with
feedback provided by the proof engine.

Figure~\ref{fig:onlineprover-protocol} illustrates how the components of
\onlineprover communicate. A user accesses the editor by visiting a URL, which
results in an HTTP GET request that responds with the exercise, along with the
sourcecode for the editor itself. The user can then begin solving the exercise
by editing the derivation tree. A user may at any time \emph{check} their
solution, which is materialized as a HTTP POST request, containing a JSON
encoding of the derivation tree. The server consults the proof engine, which
returns a derivation tree annotated with feedback. The user may continue to
edit and check their proof in a feedback-loop, similar to how a programmer
interacts with a compiler, until the proof is finally verified by the proof
engine.

The server and proof engine are written in Haskell, while the editor is
implemented in ClojureScript. Derivation tree states are stored in
\texttt{LocalStorage}, ensuring that the server and proof engine remain
completely stateless. All user interaction events, along with timestamps, are
recorded, making it possible to replay and reconstruct every state of the
user’s derivation. We also support both data export and import: exports can be
used for analysis (see Section~\ref{sec:data-analysis}), while imports enable users to
continue their work on a different machine.
 

\section{Teaching with OnlineProver: Motivating example}
\label{sect:Teaching}

In this section, we present a simple example of the use of \onlineprover from a student's perspective. For this purpose, we have deployed a version of the \onlineprover system with a set of exercises for Gentzen-style proofs using natural deduction in propositional logic and first-order logic. The exercises consist of lists of incomplete derivations, and a student must complete them in an empty context.

We have implemented exercises on natural deduction based on the notation used in \emph{The Logic Manual} \cite{halbach:2010}. This style uses an implicit context for assumptions. Therefore, when applying rules that introduce assumptions, one must name each assumption with a tag (a \textit{lowercase Latin letter}). In the graphical user interface, an exercise in this logic appears as a formula, above which the corresponding derivation must be constructed. Table \ref{tab:logicsymbols} summarizes the notation used by \onlineprover compared to the standard symbols of logic.

\begin{table}[tp]
\centering
\begin{tabular}{|c|c|}
\hline
\textbf{Logic Symbol} & \textbf{OnlineProver Notation} \\ \hline
$\forall$            & \texttt{forall}                \\ \hline
$\exists$            & \texttt{exists}                \\ \hline
$\neg$               & \texttt{$!$}                   \\ \hline
$\land$              & \texttt{$ /\backslash $}       \\ \hline
$\lor$               & \texttt{$ \backslash / $}      \\ \hline
$\rightarrow$        & \texttt{$->$}                  \\ \hline
$\bot$               & \texttt{False}                 \\ \hline
$\top$               & \texttt{True}                  \\ \hline
\end{tabular}
\caption{Mapping of Logic Symbols to OnlineProver Notation.}
\label{tab:logicsymbols}
\end{table}

\begin{figure}[tp]
    \centering
    \includegraphics[width=0.75\textwidth]{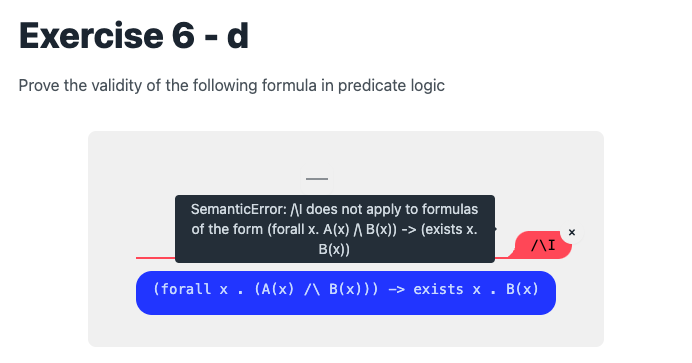}
    \caption{An example of an incorrectly applied rule.}
    \label{fig:error}
\end{figure}

As a motivating example, we prove the following formula in \onlineprover:
\begin{equation}
\label{eq:formula1}
(\forall x. (A(x) \wedge B(x))) \rightarrow \exists x . B(x)
\end{equation}
which can be translated into the \onlineprover notation as:
\begin{verbatim}
(forall x . (A(x) /\ B(x))) -> exists x . B(x)
\end{verbatim}

\begin{figure}[tp]
    \centering
    \includegraphics[width=0.75\textwidth]{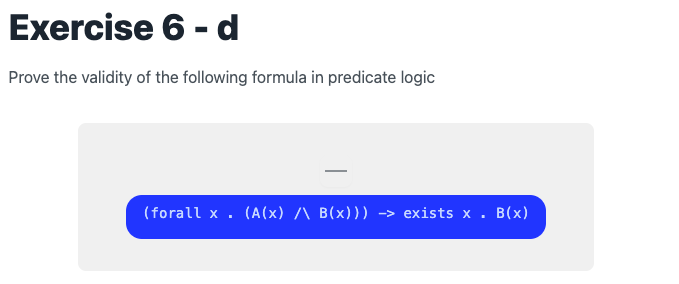}
    \caption{Exercise 6-d from \url{onlineprover.com}.}
    \label{fig:initial}
\end{figure}
Exercise 6-d in \onlineprover contains the formula (\ref{eq:formula1}), as depicted in Figure \ref{fig:initial}.
The button with a line symbol above the formula adds a new line on top of it together with the input box for the rule name. The next step is to input the rule name (\textit{with a tag for the assumption name if needed}), and, if necessary, add more input boxes for premises. \onlineprover is developed as a proof assistant, so it can check if the (\textit{partial}) derivation is correct. If not, it will provide feedback via the \textit{Check} button. For example, Figure \ref{fig:error} shows a situation where an incorrect rule was applied.

\begin{figure}[tp]
    \centering
    \includegraphics[width=0.75\textwidth]{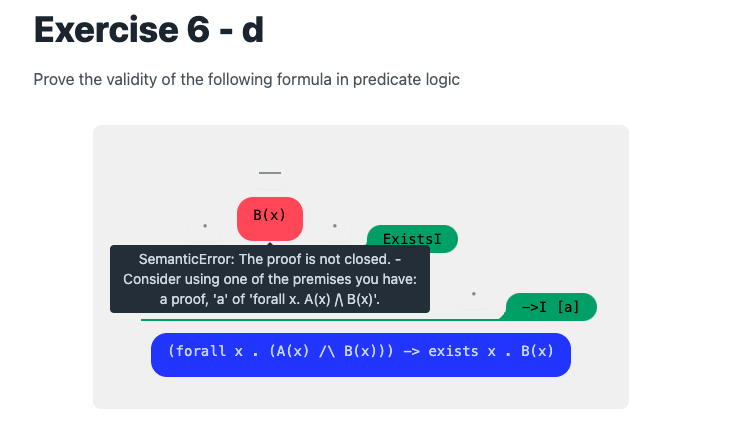}
    \caption{An example of the error with a partially written derivation.}
    \label{fig:almost}
\end{figure}

Input boxes turn red, and an error message is displayed. Once a correct (partial) derivation is provided, the correctly filled input boxes turn green. Figure \ref{fig:almost} depicts a situation where two steps of the derivation are correct, but the proof is still incomplete.

\begin{figure}[tp]
    \centering
    \includegraphics[width=0.99\textwidth]{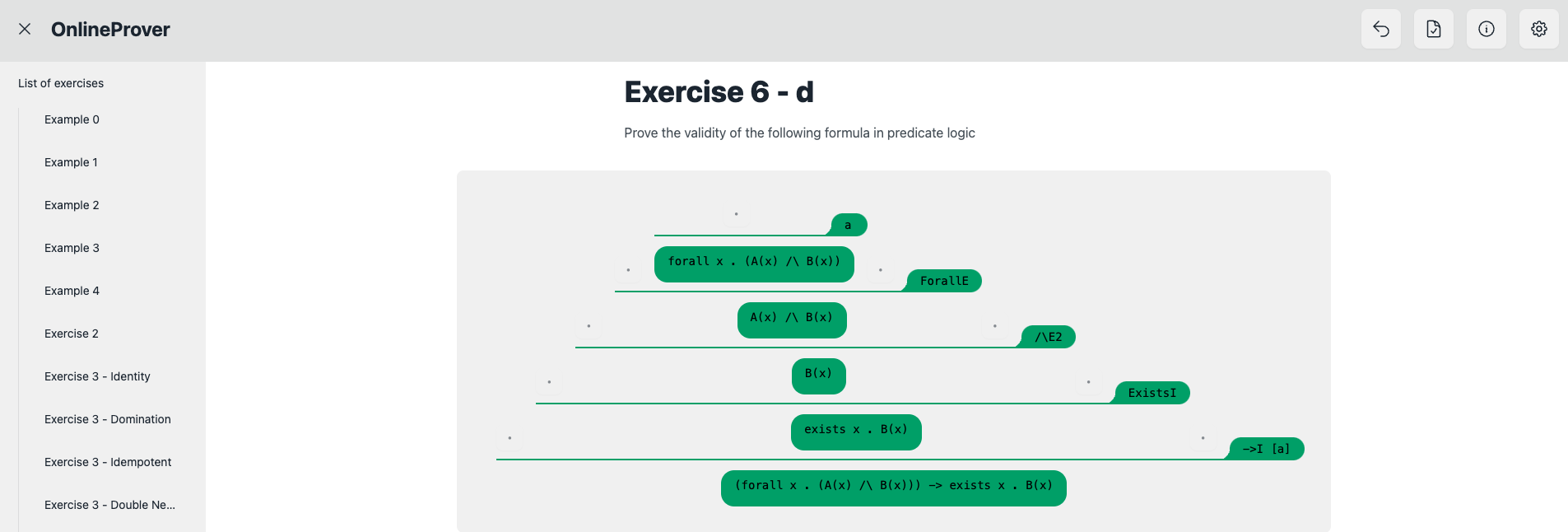}
    \caption{An examples of a closed proof.}
    \label{fig:closed}
\end{figure}

Finally, the proof is complete when all branches of the proof tree are closed by an assumption. To apply an assumption, the user must input its tag into the rule name box. The proof is considered complete when all input fields turn green. Figure \ref{fig:closed} shows the entire exercise page with the closed proof.

The top panel contains the following functionalities:
\begin{itemize}
    \item \textbf{Check button}: Provides feedback on the correctness of the derivation.
    \item \textbf{Undo button}: Reverts one step back in the derivation.
    \item \textbf{Info button}: Displays the derivation rules on the right side of the screen.
    \item \textbf{Settings button}: Allows the user to delete all derivations, change the theme, and more.
\end{itemize}


\section{Evaluation of Teaching Intervention with \onlineprover}
\label{sect:evaluation}
\label{sect:intervention}

The formal evaluation of OnlineProver is based on one semester - Autumn 2024, in the bachelor level first year course
``Foundations of Computing - Discrete Mathematics'' \cite{itucoursefoundations2024} with 165 students, at the IT University of Copenhagen.

We have decided to evaluate the deployment of \onlineprover using three methods:
\begin{itemize}
	\item The system usability scale (SUS) \cite{brooke1996sus}, which is a standardized method for evaluation of software  usability \cite{lewis2018}.
	\item We asked open questions to collect feedback on functionality, technical details, teaching, and cognitive presence.
	\item Data analysis of the EDN files provided by students.
\end{itemize}

In the rest of the section, we present the results of the SUS, questionnaire, and EDN file data analyses. The interpretation of these findings, including how they address the research questions and suggestions for improvement, will be thoroughly discussed in the next section, titled Discussion.

\subsection{Intervention Description}

As mentioned earlier, we deployed \onlineprover in the course "Foundations of Computing - Discrete Mathematics." Before \onlineprover, students used an online teaching tool called \texttt{ProofWeb}~\cite{kaliszyk2008deduction}, along with the traditional pen-and-paper approach for practicing proofs and logical reasoning. This is a first-year bachelor-level course; therefore, we can assume that students have little to no prior experience with formal proofs.

Half of the course is dedicated to propositional logic, first-order logic, and formal proofs. The first three lectures focused on logic and natural deduction. \onlineprover was introduced to students during the first lecture, where the lecturer solved example exercises from slides (referred to as \textit{"Example"} in \onlineprover). During the first lab session, students were tasked with solving problems up to and including Exercise 3-i in the current deployment. These exercises are direct transcripts of the ones at the end of each chapter in the corresponding reading material for each lecture. In the second lab session, students were asked to complete the remaining exercises up to Exercise 8-d. Students could receive help from the lecturer and teaching assistants during the lab sessions if they requested it.

During the third lecture, students were given an assignment to assess their understanding of the topic, requiring them to complete proofs in a pen-and-paper format. Students were asked to fill out the questionnaire during the same lecture, after reviewing the assignment text for the first assignment, but before submitting it for grading. After that, students were asked to export their data from \onlineprover in the EDN format and send it to us. We use the data for data analysis.

The questionnaire consisted of the SUS questions to calculate a standardized score and open questions for quantitative and qualitative evaluation.

\subsection{System Usability Scale Score}

The first part of the questionnaire consisted of the System Usability Scale (SUS) form: a standardized set of 10 questions in the Likert scale style \cite{brooke1996sus}, which is a widely used method for assessing system usability. A total of 55 students filled out the form during the third lecture of the course. Therefore, we can assume that students had sufficient exposure to the tool, allowing them to provide a meaningful assessment.

In this method, system usability is measured using a SUS score \cite{Sauro2011}, which represents a grade on a 0–100 scale, with an average score of 68.

Based on the collected data, we calculated the \textbf{SUS score} to be \textbf{67.27}. After applying a $10\%$ trimmed mean, the adjusted \textbf{score is} 67.45. This score corresponds to a grade of C.

We would like to note that the final SUS score could potentially be higher, as one of the questions in the SUS form might have been misinterpreted: "\textit{I needed to learn a lot of things before I could get going with this system.}" In this context, users might have interpreted the question as referring to their effort in learning logic and natural deduction proofs, rather than the usability of the system.

\subsection{Questionnaire}

As mentioned earlier, 55 students filled out the questionnaire in total. Since the questionnaire included both closed and open questions (and answering was optional), we divided the questions into the following categories:

\begin{description}
    \item[Quantitative questions:]~
    \begin{itemize}
        \item How many exercises did you solve on onlineprover.com?
        \item How much time would you estimate that you spent on onlineprover.com?
    \end{itemize}
    \item[Qualitative questions:]~
    \begin{itemize}
        \item Mention something that you struggled with.
        \item How would you typically resolve this particular struggle?
    \end{itemize}
    \item[Comparison (doing proofs by hand vs \onlineprover):]~
    \begin{itemize}
        \item What did you find positive about doing exercises by hand compared to using \onlineprover?
        \item What did you find positive about doing exercises using \onlineprover compared to doing them by hand?
    \end{itemize}
\end{description}

The last three questions are reserved for the discussion section:
\begin{itemize}
    \item Do you have any suggestions that would improve the quality of the tool for you?
    \item How would this improvement help you use the tool?
    \item Do you have any other suggestions?
\end{itemize}

\subsubsection{Quantitative Questions}

We asked two control questions. The first was a closed question to estimate the approximate number of exercises solved by students. We can later validate the results through data analysis of the EDN files. All 55 students answered this question.

\begin{figure}[ht!]
    \centering
    \includegraphics[width=0.8\textwidth]{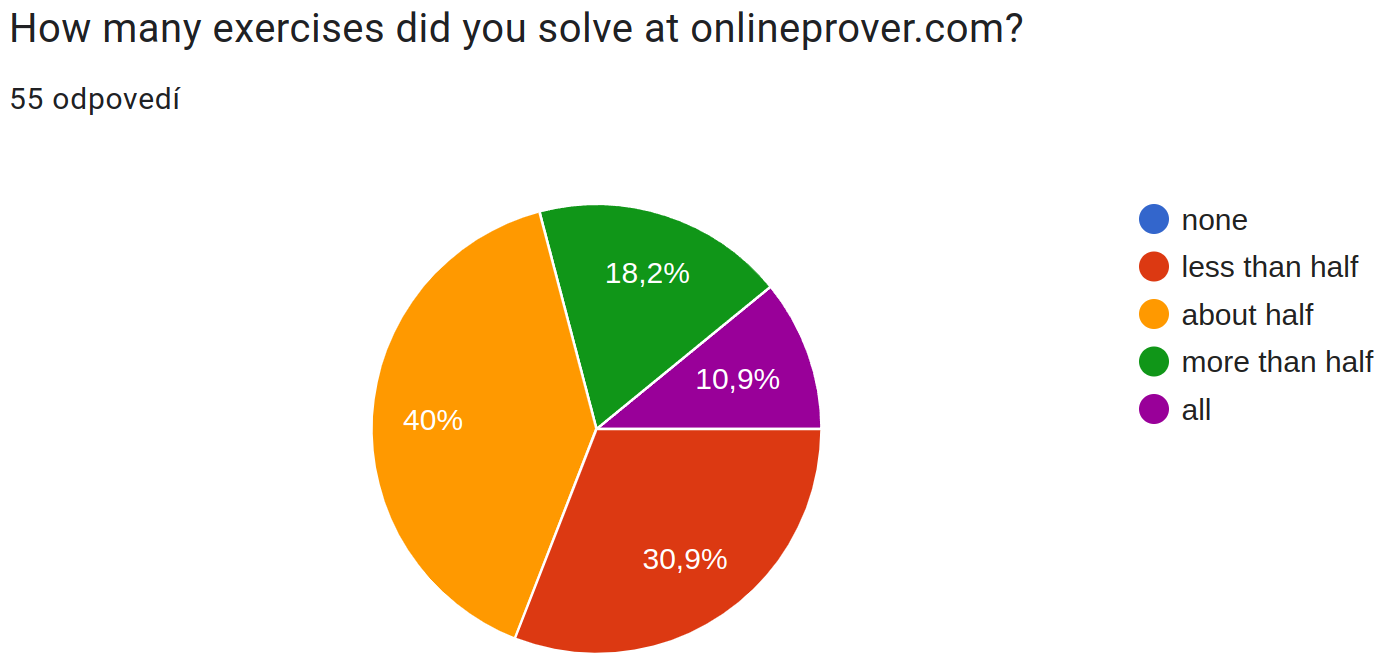}
    \caption{Question 1: Number of exercises solved on \onlineprover.}
    \label{fig:q1}
\end{figure}

The chart in Figure \ref{fig:q1} shows that $40\%$ of students claimed to have solved about half of the exercises, while approximately $30\%$ reported solving either less or more than half of the exercises. This distribution matches our expectation that the results would approximate a Gaussian distribution.

The second control question aimed to determine if students spent an appropriate amount of time using \onlineprover. This could indicate whether the tool is easy to use, well-designed, and serves its purpose effectively. We will further validate this in the discussion section based on the EDN files data analysis.

\begin{figure}[ht!]
    \centering
    \includegraphics[width=0.5\textwidth]{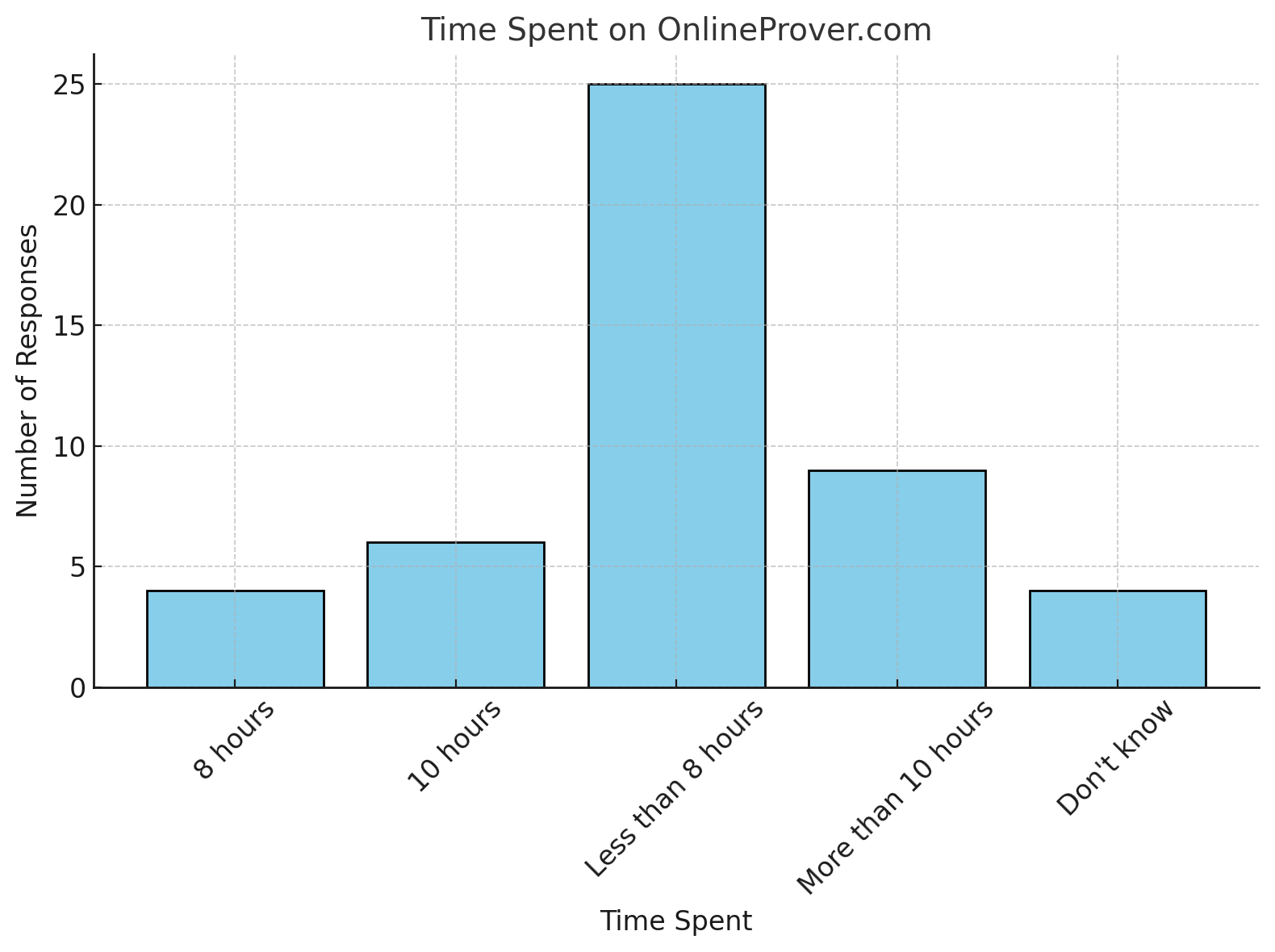}
    \caption{Question 2: Time spent on \onlineprover.}
    \label{fig:q2}
\end{figure}

We asked the following open-ended question:
\begin{quote}
\textit{How much time would you estimate that you spent on onlineprover.com?}
\end{quote}
A total of 50 students answered this question. We categorized the responses and summarized them in Figure \ref{fig:q2}.

"The most common responses were $10$ hours and $8$ hours. The largest group of students (25) reported spending less than 8 hours on \onlineprover, while 9 students reported spending more than 10 hours, with some outliers spending 20–30 hours. Additionally, one student mentioned spending more time using \onlineprover than doing pen-and-paper proofs, while another felt they should have dedicated more time to using \onlineprover."

\subsubsection{Qualitative questions}
We aimed to investigate students' behavior, particularly the types of problems they encountered and how they addressed them. To achieve this, we posed two \textit{general} open-ended questions:
\begin{quote}
1. \textit{Mention something that you struggled with.} \\
2. \textit{How would you typically resolve this particular struggle?}
\end{quote}

A total of 48 students responded to the first question, while 45 answered the second.

We analyzed the responses and categorized them based on the frequency of specific topics or issues mentioned. The challenges identified were not solely related to \onlineprover{} but also extended to broader topics, such as logic and natural deduction. Interestingly, the most common problems students faced were not with the tool itself but with understanding the theoretical concepts that the tool is designed to illustrate.

As shown in Table \ref{tab:studentproblems}, the most frequently reported challenge (20 responses) was understanding formal methods, including general comprehension of proofs and the application of deduction rules. As noted earlier, these are first-year bachelor-level students, many of whom likely have little to no prior experience with formal proofs. Other difficulties unrelated to the \onlineprover{} system included the complexity of exercises (6 responses) and the lack of a visible list of assumptions, which was mentioned once.

Among issues related to \onlineprover{}, the most common challenges involved the input system (5 responses), such as difficulties entering logical symbols in ASCII code and the inability to delete internal nodes or move entire branches in a proof tree. Additionally, 4 students reported difficulties interpreting error messages in the feedback system. Bugs were also noted by two respondents; however, these have already been corrected.

\begin{table}[ht!]
\centering
\begin{tabular}{|l|c|}
\hline
\textbf{Problem Category} & \textbf{Number of Responses} \\ \hline
\multicolumn{2}{|c|}{\textbf{Not Related to \onlineprover}} \\ \hline
Understanding deduction rules & 20 \\ \hline
Exercises are hard to solve/understand & 6 \\ \hline
Missing list of assumptions (implicit context) & 1 \\ \hline
\multicolumn{2}{|c|}{\textbf{Related to \onlineprover}} \\ \hline
Input system (difficulty entering special symbols) & 5 \\ \hline
Feedback system (hard to understand feedback) & 4 \\ \hline
Bugs (already corrected) & 2 \\ \hline
\end{tabular}
\caption{Student Problems.}
\label{tab:studentproblems}
\end{table}

As shown in Table \ref{tab:studentssolutions}, the most common strategy students used to address challenges was, as expected, asking for help from teaching assistants or classmates, with 19 responses. Trial-and-error methods were also widely used, as noted by 9 respondents, while a smaller group relied on studying materials or external sources (3 responses). Interestingly, only one respondent reported using large language models like ChatGPT as part of their problem-solving process.

\begin{table}[ht!]
\centering
\begin{tabular}{|l|c|}
\hline
\textbf{Solution Strategy} & \textbf{Number of Responses} \\ \hline
Trial and error / brute force & 9 \\ \hline
Asking a teaching assistant or a classmate for help & 19 \\ \hline
Asking large language models & 1 \\ \hline
Reading studying materials / external sources & 3 \\ \hline
\end{tabular}
\caption{Students' Solutions.}
\label{tab:studentssolutions}
\end{table}

\subsubsection{\onlineprover vs Pen and Paper Proofs}
\label{sec:comparision}

To compare the traditional approach of teaching formal proofs using pen and paper with \onlineprover, we asked the following questions:

\begin{quote}
1. What did you find positive about doing exercises by hand compared to using \onlineprover?\\
2. What did you find positive about doing exercises using \onlineprover compared to doing them by hand?
\end{quote}

These questions aim to identify student challenges and preferences, as well as compare learning efficiency and effectiveness between the two approaches.

Table \ref{tab:penandpaper} shows that the largest group of students (13 responses) considered pen and paper proofs to be more beneficial for learning and understanding. Students also appreciated the flexibility of doing proofs on paper (12 responses). Some students (3 responses) noted that pen and paper proofs allowed them to use the same notation as in the textbook, which can be particularly helpful for beginners. Additionally, 4 students mentioned that they had not completed any proofs by hand, suggesting that \onlineprover is the dominant approach in their learning process.

\begin{table}[ht!]
\centering
\begin{tabular}{|l|c|}
\hline
\textbf{Pen and Paper Proofs} & \textbf{Number of Responses} \\ \hline
Better for Learning and Understanding & 13 \\ \hline
Ease of Use and Flexibility & 12 \\ \hline
Notation like in the book & 3 \\ \hline
Have not done by hand & 4 \\ \hline
\end{tabular}
\caption{Pen and Paper Proofs Positives.}
\label{tab:penandpaper}
\end{table}

Table \ref{tab:onlineprover} shows that \onlineprover was positively received for its feedback (error messages) with 9 responses, and the possibility to check proof correctness (9 responses). The tool's user interface was also appreciated, with 7 responses mentioning that it made solving proofs faster, while 5 students appreciated its overall ease of use and flexibility. Additionally, 6 responses highlighted how \onlineprover supported their learning process.

\begin{table}[ht!]
\centering
\begin{tabular}{|l|c|}
\hline
\textbf{OnlineProver} & \textbf{Number of Responses} \\ \hline
Feedback system & 9 \\ \hline
Proof checker & 9 \\ \hline
Speed and Efficiency & 7 \\ \hline
Support for Learning & 6 \\ \hline
Ease of Use and Flexibility & 5 \\ \hline
\end{tabular}
\caption{\onlineprover Positives.}
\label{tab:onlineprover}
\end{table}

Overall, two responses stand out. One student, who is retaking the course, compared the tool ProofWeb, used in previous years, with \onlineprover and found the latter to be significantly better. This comparison highlights the improvement and effectiveness of \onlineprover in enhancing the learning experience. On the other hand, another student pointed out a potential drawback of relying solely on digital tools like \onlineprover. They noted that when switching back to pen and paper proofs, they felt insecure about their proofs, pointing out the importance of balancing the use of digital tools with traditional methods to reinforce confidence and understanding.

\subsection{Data analysis}
\label{sec:data-analysis}

We have collected the EDN files that contain history of all actions (explained in Section~\ref{sec:tooldesign}). Overall 36 students voluntarily submitted their data for evaluation. For a demonstration purposes and to validate some of our questions in the questionnaire, we have decided to analyse the student interaction with exercises, time spent, checks performed, the ratio of deletions to additions, edit operations per exercise, and the relationship between edits and checks.

%
\subsubsection{Exercise Attempts Analytics}

\begin{figure}[tp]
    \centering
    \includegraphics[width=0.9\textwidth]{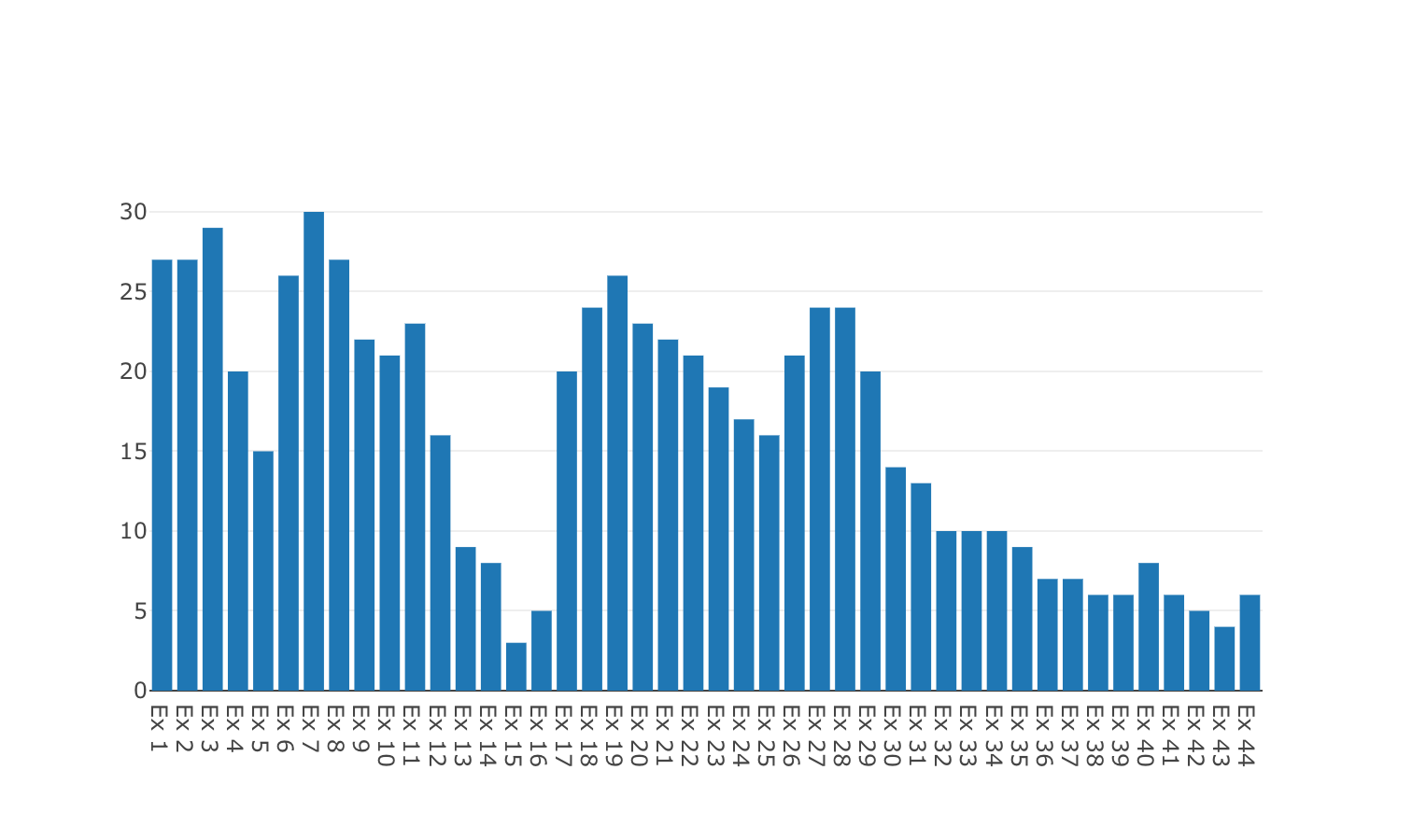}
    \caption{Number of attempts per exercise.}
    \label{fig:exatt}
\end{figure}

The current \onlineprover deployment consisted of 44 exercises. The following bar chart (Figure~\ref{fig:exatt}) shows the number of students who attempted each exercise.

From this bar chart, we can identify which exercises were less popular. As expected, the later exercises were attempted the least, but it is interesting to note that some exercises in the first half of the set (e.g., 13–17) were attempted by fewer than 10 students.

\begin{figure}[tp]
    \centering
    \includegraphics[width=0.4\textwidth]{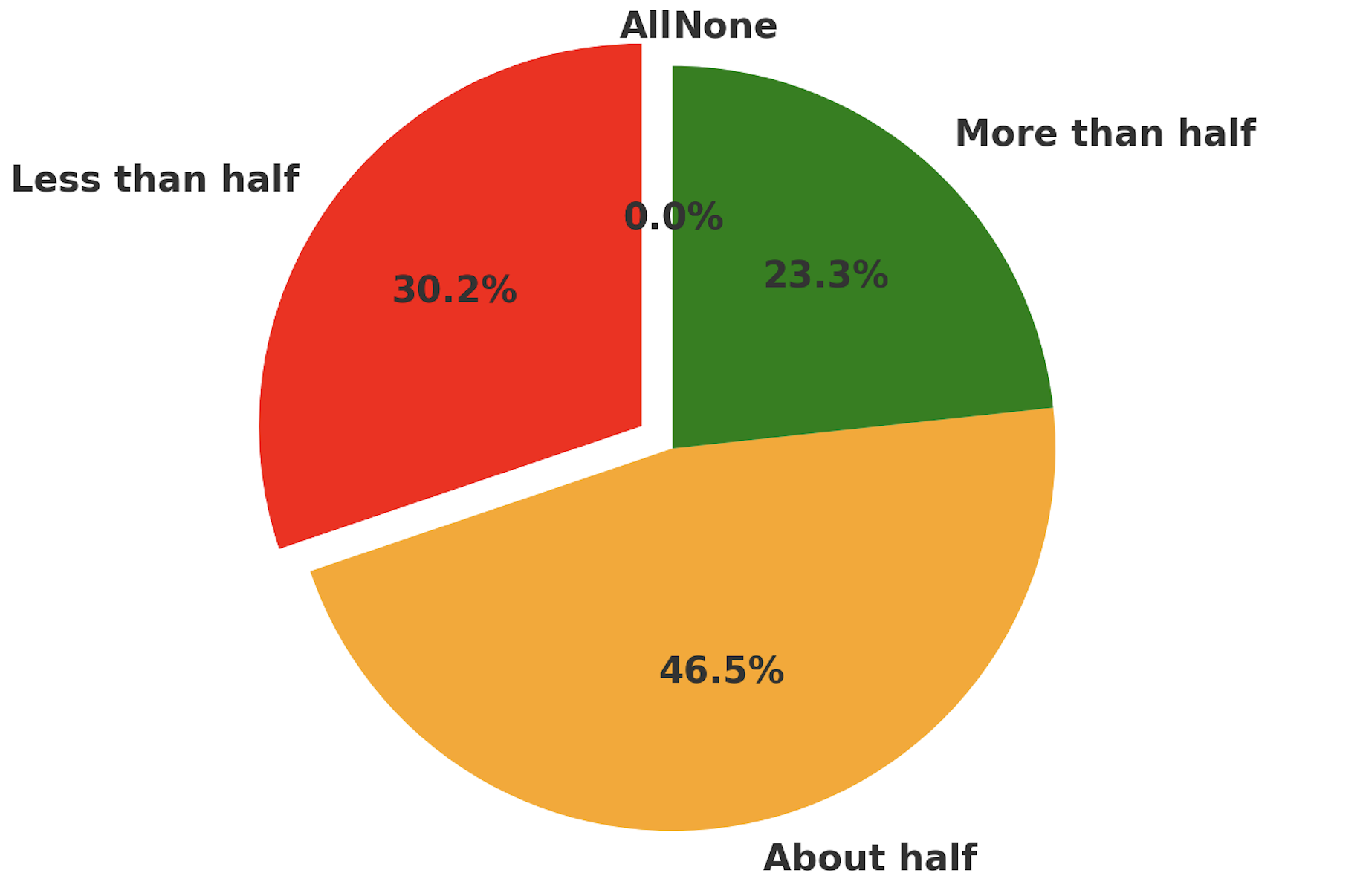}
    \caption{Exercise attempts.}
    \label{fig:exatt-pie}
\end{figure}

When this data is converted into a pie chart (Figure \ref{fig:exatt-pie}) and compared with Chart \ref{fig:q1}, similar trends emerge.

The data broadly correspond to the responses in the questionnaire, even though a different subset of students might have completed the questionnaire and submitted the EDN files. The high number of attempts in earlier exercises aligns with the larger proportion of students indicating they solved about half or fewer exercises. Only a small percentage of students solved all the exercises, which is consistent with the lower participation observed in later exercises.

\subsubsection{Time Spent Analysis}

\begin{figure}[tp] 
    \centering
    \includegraphics[width=0.99\textwidth]{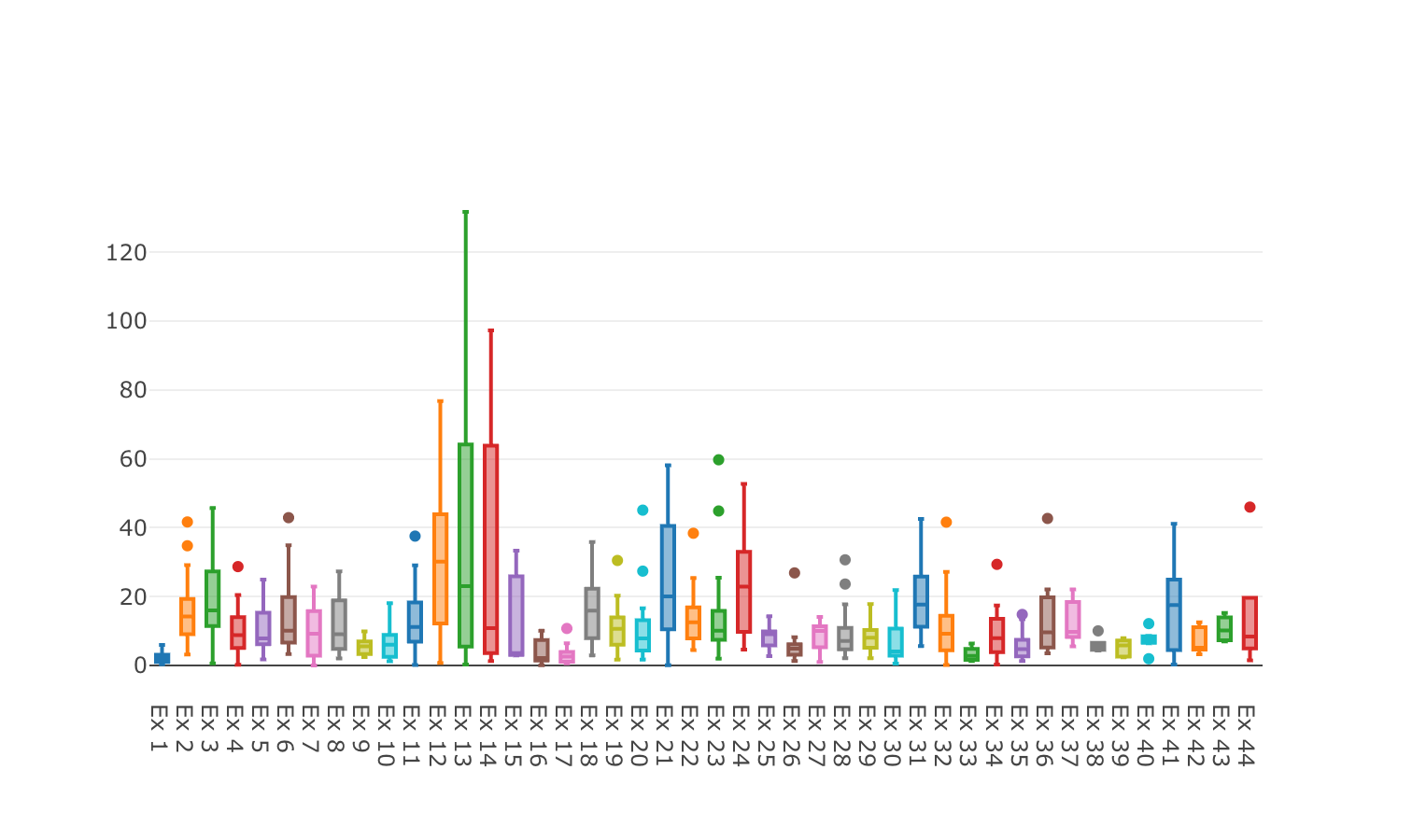}
    \caption{Time spent per exercise in minutes.}
    \label{fig:timespend}
\end{figure}

The following chart (Figure \ref{fig:timespend}) shows the average time spent per exercise in minutes.

The chart illustrates the distribution of time spent on various exercises, labeled from "EX 1" to "EX 44." It highlights significant variations in time spent across exercises, with some exercises showing outliers where participants spent considerably more time. This analysis provides insights into the relative difficulty of the exercises.

\subsubsection{Behavioural Analytics}

The following charts provide a detailed analysis of user interactions with \onlineprover, focusing on various behavioural metrics. These include the number of checks performed per exercise, the ratio of deletions to additions, the frequency of edit operations across exercises, and the relationship between edit operations and checks.

By analysing these metrics, we aim to investigate patterns in how students interact with the tool, their problem-solving approaches, the effectiveness of their trial-and-error strategies, and the usability of the system. These insights can help us understand how students solve their proofs and refine the learning process.

\begin{figure}[tp]
    \centering
    \includegraphics[width=0.99\textwidth]{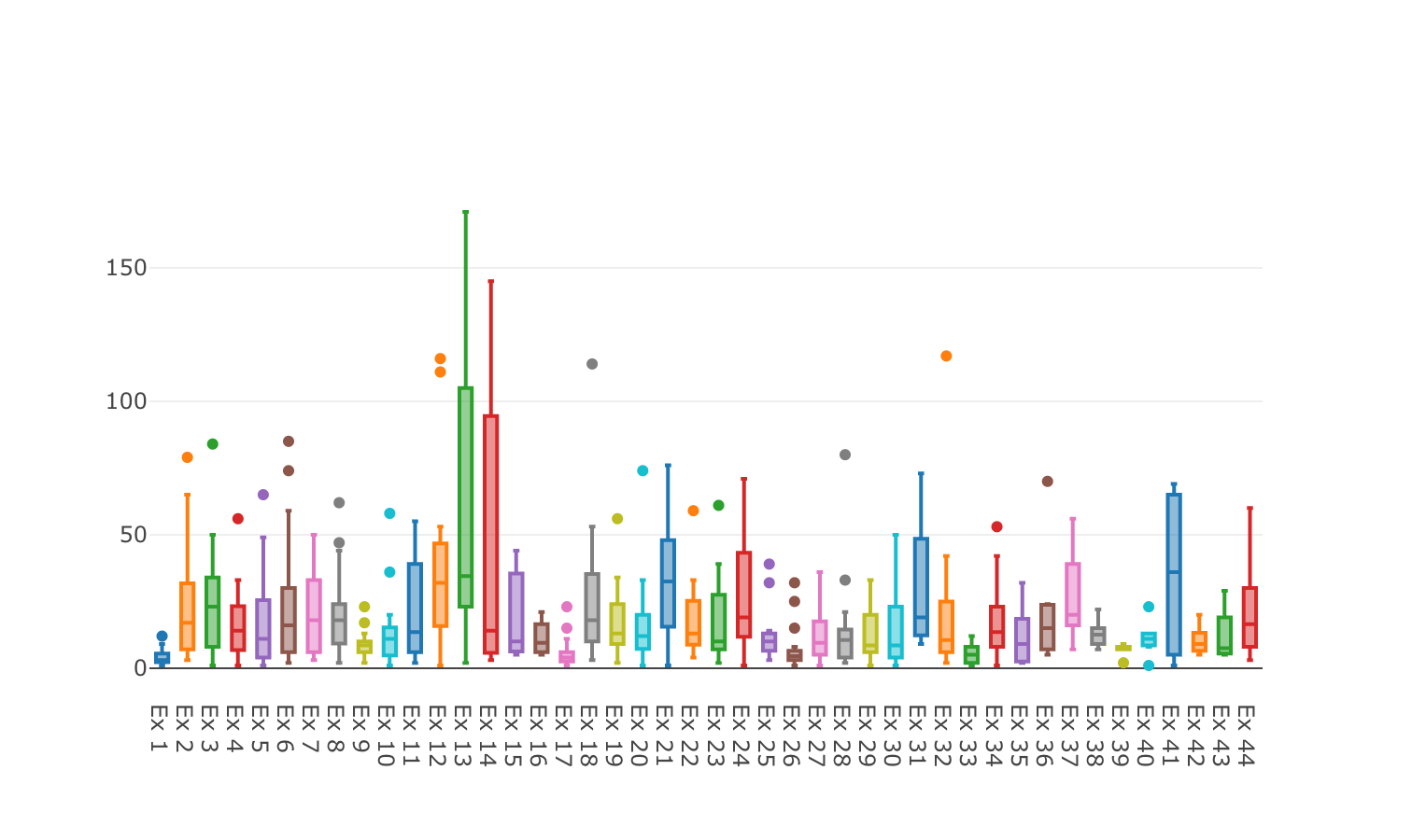}
    \caption{Number of checks per exercise.}
    \label{fig:checks}
\end{figure}

Figure~\ref{fig:checks} shows the number of times the check button was used per exercise. This feature allows students to confirm if their (partial) solution is correct or to receive feedback in case of incorrect derivations. As observed, the number of checks aligns with the time spent chart (Figure \ref{fig:timespend}).

\begin{figure}[tp].
    \centering
    \includegraphics[width=0.99\textwidth]{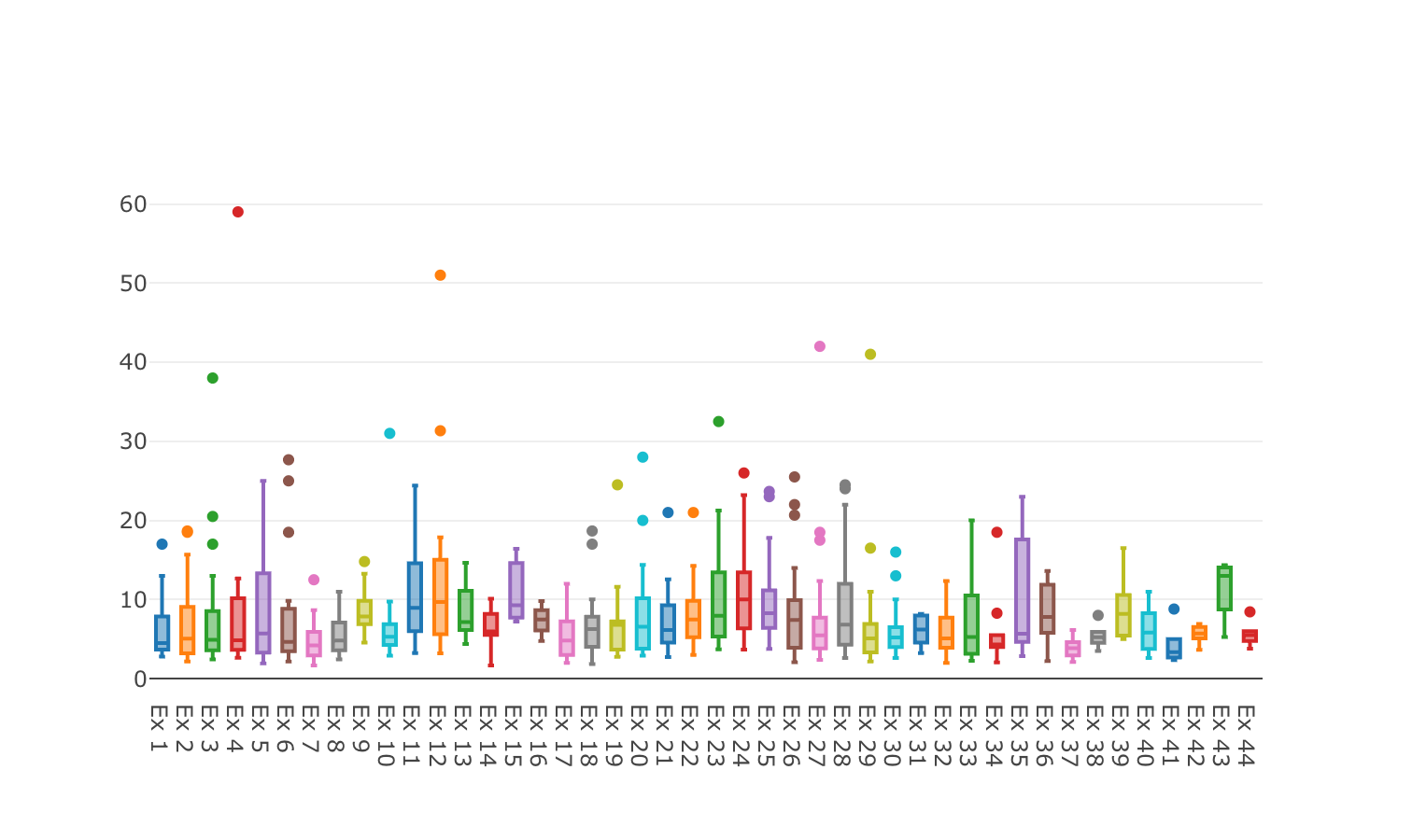}
    \caption{Ratio between edit actions and checks.}
    \label{fig:ratio}
\end{figure}

If we examine the ratio between edit actions and usage of the check feature (Figure \ref{fig:ratio}), it becomes evident that the feedback system was utilized frequently, as students, on average, used the check button at least once in every ten edit operations.

\section{Discussion}
\label{sec:discus}

In this section, we begin by discussing the suggestions for improvement and their impact collected in the questionnaire. Following that, we will address the research questions formulated in Section~\ref{sec:reseachquestion}, based on the evaluation of our intervention.

\subsection{Suggestions}

The last three questions of the questionnaire aimed to collect overall feedback on the tool, specifically how it could be improved to better meet users' needs. We were interested in understanding how students evaluate the tool, what its strengths and weaknesses are, and whether they have any additional comments or suggestions. To gather this information, we asked the following questions:

\begin{quote}
1. Do you have any suggestions that would improve the quality of the tool for you?\\
2. How would this improvement help you use the tool?\\
3. Do you have any other suggestions?
\end{quote}

\subsubsection{Suggestions for Improvement}

In the first question, we asked students for improvement suggestions. We have categorized them and ordered according the number of responses mentioning the category in Table \ref{tab:suggestions}.

\begin{table}[tp]
\centering
\begin{tabular}{|l|c|}
\hline
\textbf{Category} & \textbf{Number of Responses} \\ \hline
Formatting and Shortcuts & 11 \\ \hline
UI Suggestions & 8 \\ \hline
Ability to Create Own Proof Trees & 7 \\ \hline
Learning Support and Tutorials & 6 \\ \hline
Improve Feedback System & 2 \\ \hline
List of Hypothesis & 2 \\ \hline
\end{tabular}
\caption{Suggestions for Improvement.}
\label{tab:suggestions}
\end{table}

The feedback collected on the tool suggests possible improvement in several areas. Formatting and shortcuts being the most requested (11 responses), followed by UI suggestions (8), and the ability to create own proof trees (7).

Students mostly suggested to add a easier way for entering special symbols or keyboard shortcuts for this purpose. For that, we do plan to implement a better way of entering special symbols.  UI improvements mentioned draggable elements and customizable layouts for better user experience. The ability to create own formulae was often mentioned, including features like a sandbox mode and generating LaTeX or image outputs. Learning support received six responses, where students suggested to add detailed tutorials, rule explanations, and hints to aid understanding. Feedback system enhancements suggested that there is a room for improvement of error messages. At last 2 students mentioned that it would be nice to be able to display a list with the current hypotheses in a proof tree.

\subsubsection{Expected Impact of Improvements}

In the second question, we asked students how do they think their suggestion would improve our tool. We have categorized them and ordered according the number of responses mentioning the category in Table~\ref{tab:suggestions2}.

\begin{table}[tp]
\centering
\begin{tabular}{|l|c|}
\hline
\textbf{Category} & \textbf{Number of Responses} \\ \hline
Better Learning and Understanding & 9 \\ \hline
Time-Saving and Efficiency & 8 \\ \hline
Customization and Personalization & 3 \\ \hline
Improved Usability and Accessibility & 5 \\ \hline
Interface and Navigation & 5 \\ \hline
\end{tabular}
\caption{Expected Impact of Improvements.}
\label{tab:suggestions2}
\end{table}

The expected impact of the suggested improvements mentions several areas. Time-saving and efficiency were pointed out as major benefits, with students noting that the changes like ability to input own formulae would eliminate the need for paper, reduce redundant actions, and would make their work faster. Specific mentions included making it easier to fit large proofs on screen, requiring fewer clicks, and make the process of writing and understanding proofs easier.

Better learning and understanding was pointed out as the most important expected impact of the suggested improvement. Students mentioned the importance of using the same notation as in the textbook, smarter error messages, which would create better navigation and readability.

To address these points, we plan to implement a more intuitive system for entering special symbols, such as using \LaTeX syntax, which will automatically convert the input into the corresponding symbol. This feature will also improve readability, as special symbols will be displayed exactly as they appear in the textbook. Draggable elements and customizable layouts are under consideration to enhance usability. Additionally, we aim to introduce a sandbox mode and export options for exercises, as well as more detailed tutorials to support learning. The feedback system in \onlineprover is particularly important. We plan to expand its capabilities beyond simple error messages to a data-driven, intelligent system that provides tailored feedback based on the user's behavior.

\subsubsection{Additional Comments}

In the last question, we asked students if they want to add any other comment. We have categorized answers in Table~\ref{tab:suggestions3}.

\begin{table}[tp]
\centering
\begin{tabular}{|l|c|}
\hline
\textbf{Category} & \textbf{Number of Responses} \\ \hline
Customization and Personalization & 3 \\ \hline
Improved Usability and Accessibility & 6 \\ \hline
Interface and Navigation & 4 \\ \hline
Own proof tree & 4 \\ \hline
\end{tabular}
\caption{Other Comments.}
\label{tab:suggestions3}
\end{table}

Other suggestions focused on customization, usability, and interface improvements to enhance the flexibility and user experience of \onlineprover. Students requested customizable themes and options to reduce reliance on shift-key inputs for typing. Other suggestions included the ability to save progress and make the layout more compact to handle wide proofs effectively. For interface and navigation, recommendations included clearer distinctions between predicate and propositional logic and smoother transitions between exercises. Additionally, students requested the ability to enter their own formulae.

To address these suggestions, we plan to implement the following. As mentioned earlier, we will introduce a new input system. The ability to change themes has already been implemented, and we will incorporate this feature into the plan tutorials. Saving progress and importing it back is another feature we plan to implement in future versions. Regarding navigation, we will improve the distinction between topics and make switching between them easier.

\subsection{Addressing the Research Questions}

In this section, we address the research questions formulated in the Section~\ref{sec:reseachquestion}.

\subsubsection{RQ1}

Our first research question is as follows:

\vspace{-1mm}
\begin{quote}
Is our tool used as a proof assistant or a proof checker?
\end{quote}
\vspace{-1mm}

\noindent
As discussed in the Evaluation and Discussion sections, the feedback system is frequently mentioned by students, even though we did not ask about it directly. From Table~\ref{tab:onlineprover}, it is evident that students positively highlighted the feedback system and the check feature, each mentioned exactly nine times. Based on this, we can conclude that \onlineprover is used both as a proof assistant and as a proof checker. 

\subsubsection{RQ2}

The feedback system was designed to simulate the type of guidance a teacher would provide—offering hints rather than solving problems outright—so that students could independently work through their proofs. While many online tools for reasoning provide solutions directly (similar to a calculator), our goal was to develop a system that helps students solve proofs on their own. Since the tool targets computer science students, we anticipated they would adopt a trial-and-error strategy similar to programming. This led us to formulate the following question:

\begin{quote}
How do students use trial-and-error strategies with \onlineprover to overcome difficulties in solving proofs?
\end{quote}

\noindent
As shown in Table \ref{tab:studentssolutions}, the trial-and-error approach is the second most common strategy used for problem-solving. This hypothesis is further supported by the analysis of EDN files, as shown in Figure~\ref{fig:ratio}, where students, on average, used the check button after almost every derivation step.

\subsubsection{RQ3}

We spent a lot of time designing and developing a useful feedback system. As mentioned earlier, we aimed to strike a balance between providing enough information to give a hint on how to continue and explaining where the derivation is incorrect, without turning our tool into a "calculator." This led us to ask the following question:

\begin{quote}
Do students perceive the tool’s feedback as useful?
\end{quote}

\noindent
The data analysis and students' responses in the questionnaire confirm our hypothesis. Although four students reported that the error messages could be improved, overall, we can conclude that the tool's feedback system was positively evaluated.

\subsubsection{RQ4}

The last question aimed to investigate whether students consider our tool more useful for learning formal proofs compared to the traditional pen-and-paper approach. We consider our tool to be a teaching aid; it should not replace pen-and-paper proofs but rather supplement them. Therefore, we formulated the following question:

\begin{quote}
Do students consider learning proofs using OnlineProver more effective than learning proofs by pen and paper?
\end{quote}

\noindent
When analyzing the responses in the comparison section (Section~\ref{sec:comparision}), we found that students consider pen-and-paper proofs to be more effective for learning than using the tool. However, they also valued \onlineprover positively as a teaching aid, particularly the feedback system and check feature, which enhanced the learning process. Thus, our initial hypothesis can be considered confirmed.

%
%
%
%
%
%
%
%
%
%
%
%
%
%


\section{Conclusion and Future Work}
\label{sect:conclusion}
\onlineprover has demonstrated significant potential as a teaching tool for enhancing the understanding of formal proofs in computer science education. Its user-friendly design, together with a realtime feedback system, helps students address challenges in learning abstract logical concepts. The System Usability Scale (SUS) score and student feedback indicate that while the tool is well-received, there is room for improvement, particularly in usability features and feedback clarity.

The trial-and-error strategy supported by \onlineprover encourages active learning and aligns well with the pedagogical needs of computer science students. The integration of structured error messages enhances reasoning with proofs, offering a effective learning experience. Notably, while students viewed the tool positively, they also emphasized the importance of traditional methods, such as the use of pen-and-paper proofs.
\onlineprover is an innovative tool offering valuable features for both teachers and students. For teachers, it enables analysis of student behavior, helping identify ``problematic'' exercises and pinpointing areas where students struggle. From a student’s perspective, \onlineprover serves as a practical learning aid without requiring the mastery of a new language, as would be necessary with full-scale proof assistants. Students benefit from real-time feedback and the ability to check correctness of their (partial) solutions.

Future work will focus on improving the tool's interface, enhancing the feedback system, and adding more interactive tutorials. The current deployment relies on a hard-coded proof engine. A significant future challenge is the planned development of two domain-specific languages: one for defining formal methods and their deduction rules, and another for creating exercises. These will compile into a proof engine, expanding the tool's flexibility and scope.


\bibliographystyle{eptcs}
\bibliography{bibliography}
\end{document}